\documentstyle[12pt]{article}

\catcode`\@=11
\long\def\@makefntext#1{
\protect\noindent \hbox to 3.2pt {\hskip-.9pt
$^{{\ninerm\@thefnmark}}$\hfil}#1\hfill}		

 \def\@makefnmark{\hbox to 0pt{$^{\@thefnmark}$\hss}}  

\def\ps@myheadings{\let\@mkboth\@gobbletwo
\def\@oddhead{\hbox{}
\rightmark\hfil\ninerm\thepage}
\def\@oddfoot{}\def\@evenhead{\ninerm\thepage\hfil
\leftmark\hbox{}}\def\@evenfoot{}
\def\sectionmark##1{}\def\subsectionmark##1{}}

\newcounter{sectionc}\newcounter{subsectionc}\newcounter{subsubsectionc}
\renewcommand{\section}[1] {\vspace{0.6cm}\addtocounter{sectionc}{1}
\setcounter{subsectionc}{0}\setcounter{subsubsectionc}{0}\noindent
	{\bf\thesectionc. #1}\par\vspace{0.4cm}}
\renewcommand{\subsection}[1] {\vspace{0.6cm}\addtocounter{subsectionc}{1}
	\setcounter{subsubsectionc}{0}\noindent
	{\it\thesectionc.\thesubsectionc. #1}\par\vspace{0.4cm}}
\renewcommand{\subsubsection}[1]
{\vspace{0.6cm}\addtocounter{subsubsectionc}{1}
	\noindent {\rm\thesectionc.\thesubsectionc.\thesubsubsectionc.
	#1}\par\vspace{0.4cm}}

\newcounter{appendixc}
\newcounter{subappendixc}[appendixc]
\newcounter{subsubappendixc}[subappendixc]

\renewcommand{\appendix}[1] {\vspace{0.6cm}
        \refstepcounter{appendixc}
        \setcounter{figure}{0}
        \setcounter{table}{0}
        \setcounter{equation}{0}
        \renewcommand{\thefigure}{\Alph{appendixc}.\arabic{figure}}
        \renewcommand{\thetable}{\Alph{appendixc}.\arabic{table}}
        \renewcommand{\theappendixc}{\Alph{appendixc}}
        \renewcommand{\theequation}{\Alph{appendixc}.\arabic{equation}}
        \noindent{\bf Appendix \theappendixc #1}\par\vspace{0.4cm}}

\def\abstracts#1{{
	\centering{\begin{minipage}{30pc}\tenrm\baselineskip=12pt\noindent
	\centerline{\tenrm ABSTRACT}\vspace{0.3cm}
	\parindent=0pt #1
	\end{minipage}}\par}}

\renewenvironment{thebibliography}[1]
	{\begin{list}{\arabic{enumi}.}
	{\usecounter{enumi}\setlength{\parsep}{0pt}
\setlength{\leftmargin 1.25cm}{\rightmargin 0pt}
	 \setlength{\itemsep}{0pt} \settowidth
	{\labelwidth}{#1.}\sloppy}}{\end{list}}

\newcounter{itemlistc}
\newcounter{romanlistc}
\newcounter{alphlistc}
\newcounter{arabiclistc}

\newcommand{\fcaption}[1]{
        \refstepcounter{figure}
        \setbox\@tempboxa = \hbox{\tenrm Fig.~\thefigure. #1}
        \ifdim \wd\@tempboxa > 6in
           {\begin{center}
        \parbox{6in}{\tenrm\baselineskip=12pt Fig.~\thefigure. #1}
            \end{center}}
        \else
             {\begin{center}
             {\tenrm Fig.~\thefigure. #1}
              \end{center}}
        \fi}

\newcommand{\tcaption}[1]{
        \refstepcounter{table}
        \setbox\@tempboxa = \hbox{\tenrm Table~\thetable. #1}
        \ifdim \wd\@tempboxa > 6in
           {\begin{center}
        \parbox{6in}{\tenrm\baselineskip=12pt Table~\thetable. #1}
            \end{center}}
        \else
             {\begin{center}
             {\tenrm Table~\thetable. #1}
              \end{center}}
        \fi}

\def\@citex[#1]#2{\if@filesw\immediate\write\@auxout
	{\string\citation{#2}}\fi
\def\@citea{}\@cite{\@for\@citeb:=#2\do
	{\@citea\def\@citea{,}\@ifundefined
	{b@\@citeb}{{\bf ?}\@warning
	{Citation `\@citeb' on page \thepage \space undefined}}
	{\csname b@\@citeb\endcsname}}}{#1}}

\newif\if@cghi
\def\cite{\@cghitrue\@ifnextchar [{\@tempswatrue
	\@citex}{\@tempswafalse\@citex[]}}
\def\citelow{\@cghifalse\@ifnextchar [{\@tempswatrue
	\@citex}{\@tempswafalse\@citex[]}}
\def\@cite#1#2{{$\null^{#1}$\if@tempswa\typeout
	{IJCGA warning: optional citation argument
	ignored: `#2'} \fi}}

\def\fnt#1#2{\footnotetext{\kern-.3em
	{$^{\mbox{\sevenrm #1}}$}{#2}}}

 1
 1
 1

\font\tenbf=cmbx10
\font\tenrm=cmr10
\font\tenit=cmti10

\font\ninerm=cmr9

\newcommand{\be}{\begin{equation}}
\newcommand{\ee}{\end{equation}}
\newcommand{\bea}{\begin{eqnarray}}
\newcommand{\eea}{\end{eqnarray}}

\begin{document}
\begin{flushright}
hep-ph/9412234 \\
LA-UR-94-3953;~~LAEFF-94/14 \\
{}~~~~ \\
\end{flushright}

\centerline{\tenbf THEORETICAL MOTIVATION FOR GRAVITATION EXPERIMENTS  }
\baselineskip=16pt
\centerline{\tenbf ON ULTRA-LOW ENERGY
ANTIPROTONS AND ANTIHYDROGEN}
\vspace{0.8cm}
\centerline{\tenrm MICHAEL MARTIN NIETO and T. GOLDMAN\footnote{Email:
mmn@pion.lanl.gov; goldman@hotelcal.lanl.gov}}
\baselineskip=13pt
\centerline{\tenit Theoretical Division}
\baselineskip=12pt
\centerline{\tenit Los Alamos National Laboratory}
\baselineskip=12pt
\centerline{\tenit University of California}
\baselineskip=12pt
\centerline{\tenit Los Alamos, New Mexico 87545 USA}
\vspace{0.9cm}
\centerline{\tenrm JOHN D. ANDERSON and EUNICE L. LAU\footnote{Email:
JDA@grouch.jpl.nasa.gov; eunice@helen.jpl.nasa.gov}}
\baselineskip=13pt
\centerline{\tenit Jet Propulsion Laboratory}
\baselineskip=12pt
\centerline{\tenit California Institute of Technology}
\baselineskip=12pt
\centerline{\tenit Pasadena, California 91109 USA}
\vspace{0.9cm}
\centerline{\tenrm J. P\'EREZ-MERCADER\footnote{Email:
mercader@estrella.laeff.esa.es}}
\baselineskip=13pt
\centerline{\tenit Laboratorio de Astrof\'isica Espacial y F\'isica
Fundamental}
\centerline{\tenit Aptdo. 50727}
\baselineskip=12pt
\centerline{\tenit 28080 Madrid, Spain}


\vspace{0.9cm}
\abstracts{We know that the generally accepted theories of gravity and quantum
mechanics are fundamentally incompatible.  Thus, when we try to
combine these theories, we must beware of physical
pitfalls. Modern theories of quantum gravity are trying to overcome
these problems.  Any ideas must confront the present agreement with
general relativity, but yet be free to wonder about not understood
phenomena, such as the dark matter problem and the
anomalous spacecraft data which we announce here.
This all has led some
``intrepid" theorists to consider a new gravitational regime, that of
antimatter.  Even more ``daring" experimentalists are attempting, or
considering attempting,  the measurement of the gravitational force
on antimatter, including low-energy antiprotons and, perhaps most
enticing, antihydrogen.
}
\vfil
\rm\baselineskip=14pt


\section{Introduction}

Classical, worldline, general relativity, and many-path quantum mechanics, are,
by the descriptive words, worldline vs. many-path, fundamentally in conflict
with each other.  It makes no difference if this distinction
would manifest itself
only at the Planck scale. It is still there.

Indeed, much of the effort of modern theoretical physics is devoted to
overcoming this conflict, at least in principle.  On the one hand, theories of
quantum gravity try to incorporate a gravitational interaction, albeit in some
higher symmetry or space-time, and then have general relativity fall out as a
classical approximation.  On the other hand, cosmologists of the ``wave
function of the universe" school, are trying to modify quantum mechanics to
allow a unified picture of physics.

Independent of whether either of these schools is on the right track, it is
logically clear that either 1) general relativity, or 2) quantum mechanics, or
3) both theories have to be modified to obtain a better theory of physics.
Ironically, our guide may be in looking at the great body of experimental data
which defines the successes of these theories:  from the successes of QED
\cite{kinoshita} to the
successes of general relativity.\cite{will,willplus}

It is as a devil's advocate  that I (MMN)
 discuss the situation for this present collaboration.  I will argue that
in many regimes we know much less than we generally assume we know.   This
opens
up the possibility that there may be something totally unexpected waiting for
us when we ultimately reach the essential
confrontation of these two fields, in
the gravity of antimatter.\cite{ng}


\section{What Do We Not Know?}

Given the many successes of general relativity, then, why would one even
question that gravity  on antimatter might be different than
gravity on matter?
To begin, one can give a two-fold rationale.    The first is exemplified in
Aspect's experimental test\cite{aspect} of Bell's inequalities.\cite{belltest}
The result (that quantum mechanics is correct) was ``known" before the
experiment.  But yet, it was important to do the experiment.  Secondly, even
in areas
where one already has an answer to a known accuracy, it is important to
significantly improve the experimental agreement.  That was emphasized by
Dicke,\cite{dicke} who argued, ``It is clear \dots that if one believes that
general relativity is established beyond question by its elegance, beauty, and
the three famous experimental checks, then the E\"otv\"os experiment has
no point! \dots  However,  if gravitational theory is to be based on
experiment,
\dots"  And so, Dicke did  his experiment.

\subsection{Gravity and CPT}

Furthermore, and as indicated in the introduction, modern attempts to unify
gravity with the other forces lead to the generic conclusion that $g(p) \neq
g(\bar{p})$, at {\it some} level.  Now this statement does not contradict CPT,
even though one might have thought so.

CPT tells us that an antiapple falls to an antiEarth in the same way that an
apple falls to the Earth.  It says nothing about how an antiapple falls to the
Earth.

{\it  But I already have cheated on you.  No
general CPT Theorem has been proven for
curved space-time general relativity.}  In fact, in some string theories,
CPT is known to be violated.   It is OK to use
CPT for intuitive arguments, but not for precise,
general arguments.  That is to say, one can expect that the statements about
apples and antiapples given above are approximately correct, but one has to be
careful if statements are given about orbiting black holes vs orbiting
antiblack
holes.

What else do we really not understand?

\subsection{CP violation}

We really don't understand CP violation as manifested in the
$K_0-{\bar{K}}_0$
system.  Remember, a parametrization, the CKM-matrix, is {\it not} an
explanation.  Furthermore, upon the existence of CP violation
 depends our supposed
understanding of the dominance of matter over antimatter in the universe.

Recall that, since the early days, some have suggested that there is a
connection between the neutral-$K$ system and gravity.\cite{good,bell}.
More recently some string theories have found CPT and CP
violation,\cite{ellis,KosPot} although,
from experiment, the amount allowed by the first of
these theories\cite{ellis} has been shown to be small.\cite{peskin}

Even more interesting is the unusual suggestion of Chardin,\cite{chardin} that
CP violation is a reflection of a microscopic violation of the arrow of time,
that is, antigravity.  What is lost is the permanence of matter.  There is a
Hawking-like radiation with a connection to entropy.

What else don't we understand about gravity?


\section{Actually, We Don't Understand Gravity and Matter
(let alone antimatter)
for Almost All of the Universe!!}

This is the dark matter\footnote{More generally one  might  use the term ``dark
mass" or the ``mass-energy" of Wheeler.}
problem, the presentation and possible resolutions
of which can be traced to the beginnings of the last century.  What does one
think if one observes
 an object that is behaving ``incorrectly" from a gravitational point of
view?  Either 1) there is unseen matter causing the odd motion, or 2) there is
a
breakdown in Newton's Law.  When the orbit of Uranus was found to be behaving
strangely 150 years ago, both John Couch Adams and Urbain Jean Joseph Leverrier
 decided that there had to be a new, unseen planet causing the
perturbations \dots and indeed there was, Neptune.\cite{grosser}   But just ten
years later, a new planet, ``Vulcan," was not the cause of the anomalous
advance
of Mercury's perihelion.  It took another 50 years for this explanation to
come, the breakdown of Newton's Law  in general relativity.

\subsection{Dark matter and large-distance scales}

Such is the problem today, on the grandest scales of the universe.  One can
observe beautiful gravitational lensing of distant galaxies by foreground
clusters of galaxies.\cite{tyson}  But the  visible matter in the
clusters is only a small fraction of what would be needed to have lensed the
distant galaxy.  Either there is dark matter in the clusters or else the
interaction  causing the lensing (it does not have to be general
relativity) is stronger than believed.\footnote{This leaves aside interesting
and complicating issues such as the controversial, new values given for the
Hubble Constant.}
No one knows.  A lot of people think,
but no one knows.

On smaller, but still long-distance scales, there are the puzzling rotation
curves of galaxies.  By Doppler measurements one can find the velocity of stars
in spiral galaxies as a function of their distance from the galactic centers.
Over a large variation of distance, this velocity is often approximately
constant.  But when one tries to account for such a velocity distribution
from  the
visible matter in these galaxies, there is not enough visible matter to account
for the motions.  Therefore, one normally presumes that there is dark matter
in the galaxies.

However, it has been observed by a number of people that certain non-Newtonian
potentials could account for the motions using the visible matter
alone.\cite{tremain,mann,milgrom}  These calculations are not precise,
since visible mass determination is also not precise. but the results are
intriguing. I will go over one of them, the Modified
Newtonian Dynamics (MOND) of Milgrom and Bekenstein.\cite{milgrom}
This model is controversial, but it serves as
a good reference point for what comes later.

Basically, this dynamics  comes from a model
equation of the form
\be
\mu(g/a_0){\bf g} = {\bf g_N}~,
\ee
where $g_N$ is the Newtonian acceleration, $g$ is the true
acceleration, $\mu$ is a monotonic function that satisfies
\be
\mu(x) \rightarrow
\left\{ \begin{array}{ll}
       x~,  ~~~~  &  \mbox{ $ x \rightarrow 0 $} \\
                 ~~~ & ~~~ \\
         1~, ~~~~   &  \mbox{ $x\rightarrow \infty$ ~}  \\
         \end{array}   \right. ~.   \label{mu}
\ee
$a_0$ is a new, critical acceleration constant, that I will
return to quickly.   The idea is that
you have a Newtonian force for large accelerations and a $1/r$
force for small accelerations.  Specifically, one  has
\be
g = \left\{ \begin{array}{ll}
       {GM}/{r^2} ~,
{}~~~~  &  \mbox{ $ g~\gg~a_0 $} \\
                 ~~~ & ~~~ \\
         \left[GM a_0\right]^{1/2}/{r}~,
 ~~~~   &  \mbox{ $g~\ll~a_0$ ~}  \\
         \end{array}   \right. ~.
\ee

$a_0$ is proportional to the Hubble-Constant-squared.
But for this constant equal to 100 (in the usual units), the value
of $a_0$ is
\be
a_0 = (2-8) \times 10^{-10}~(m/sec^2)~.
\ee
This new force allows
many galactic-rotation curves to be
explained.\cite{milfrance}

\subsection{Astronomical-Unit scales}

The distance scale at which the Sun's Newtonian force  would equal its
$1/r$ MOND force is a few thousand Astronomical Units.  One might
hope to  find corrections at smaller distances than this. This fact
suggests one look for such deviations from Newton's Law at the many AU scale,
no matter what the origin might be.
Indeed, in so doing we are simply following the suggestion of
Poincar\'e:\cite{poincare}
``\dots the true aim of celestial mechanics is not to calculate
the ephemerides, because for this purpose we could be satisfied with
a short-term forecast, but to ascertain whether Newton's law is sufficient to
explain all the phenomena."
(Note that this is in a different regime
than the much shorter laboratory and geophysical scales which have recently
been the object of much study.\cite{will2,eph})

The first place might be double stars.  These objects have been know for
about 200 years.  The problem is to track their orbits.  ``Long-period
binaries" are not even known for certain to be bound.
That they travel together is
what we know about them.  Even the orbits of shorter-period binaries have not
been studied extensively enough to look for deviations from Newtonian dynamics.
This may be a problem for further investigation.  (Note, also, that this is a
different regime than the well-studied binary pulsars or even close,
relativistic binaries \cite{binary}.  The present study deals with weak-field
systems.)

When it comes to comets, the longest-period repeating comet is the comet
discovered by Caroline Herschell in 1788 and rediscovered by Rigollet in
1939.\cite{rigo}    It goes out to a distance of 57 AU from the Sun.  However,
because of perturbations from the major planets and loss of mass in its orbit,
all this comet can tell us is that Newton's Law is approximately correct, say
to
a few percent, out to such distances. (For example, its calculated orbital
period is 155 years vs. the observed 151.)


\section{Spacecraft Tests of Gravity}

Better limits on Newtonian gravity
on these scales can be obtained with the data
from deep space probes.

\subsection{Astronomical-Unit scales}

For example,  Pioneers 10 and 11 were launched in 1972, and were the
first close encounters with the major planets, most specifically Jupiter.
After Pioneer 10's encounter with
Jupiter, and Pioneer 11's encounter with Saturn,
they eventually went into orbits in opposite directions from each
other, near the ecliptic.\footnote{After Saturn flyby, Pioneer 11 was
inclined to the ecliptic and, at
the end of Doppler tracking in August, 1990, was at ecliptic
latitude 16 arcdegrees.}
  They are in a gyro mode (rotating every 13 seconds), so their
motions are not disturbed by attitude-control thrusters, as in
the case of the Voyager 1 and 2 spacecraft.  The Pioneers'
velocities have been monitored by the NASA/JPL Deep Space Network
(DSN) using transponded coherent radio Doppler data (13 cm
wavelength) referenced to hydrogen-maser clocks at stations in
California, Australia, and Spain.  These data exist out to 30 AU
for Pioneer 11 and have been analyzed out to 57 AU for Pioneer
10.  The latter spacecraft is still returning high-quality
coherent Doppler data at 61 AU distance, and additional data
analysis is underway.

Preliminary analysis by the JPL team
indicates a systematic deviation from Newtonian
dynamics.  In order to fit the Doppler data from both Pioneer
spacecraft, they
require an excess acceleration of $8 \times 10^{-10}$ m s$^{-2}$ directed
toward the Sun.\footnote{The similarity of this number to the MOND critical
acceleration is amusing, if not intriguing.}
Although some of this excess could be explained by
nonisotropic thermal emission, it is very difficult to account
for all of it that way.

A similar (but possibly smaller) constant acceleration has been
observed on the Galileo spacecraft during its cruise trajectory
between Earth and Jupiter.  However, the Galileo excess
acceleration could be caused by a nonisotropic thermal component
of about 200 W.  Knowing that about 500 W is being delivered to
the spacecraft bus by Radioactive Thermoelectric Generators
(RTG), the JPL team   would find  this to be a remarkably large nonisotropic
component, but still plausible.

Further, the JPL people   are currently analyzing data
from the Ulysses spacecraft during its out-of-ecliptic journey
from 5.3 AU, near Jupiter's orbital radius, to its perihelion distance
at 1.3 AU.  So far it seems that a  constant acceleration,
similar to that acting on Pioneer and Galileo, is also acting on Ulysses.
But they need more data analysis to be sure.

Let me  emphasize again that all these results and conclusions
are preliminary.
When all their analysis is
complete, the JPL team  will publish the
details and  their final conclusions.
Quite properly, the JPL scientists are devoting much
 effort  into searching for a
nongravitational origin of their systematics.

Also, another problem they are considering is how
planetary orbits would be affected
by a constant radial acceleration of the magnitude indicated by the
spacecraft.  A preliminary analysis indicates that
systematic error in the orbits of the outer three major planets
could easily mask the constant acceleration.  Only Jupiter, with
its eleven year sidereal period and with spacecraft fixes on its
orbital motion, might reveal a constant acceleration
of this magnitude or, on the
other hand, rule it out at a 5.2 AU distance.  Radio ranging data
generated with the Martian Viking Landers probably rule out a
constant acceleration from the Sun
on the planets at Earth distance (1.0 AU)
and Mars distance (1.52 AU).

Weak limits on Newtonian gravity have previously been
set on
 the outer planets by means of searches for dark matter
\cite{and89,and94} and a
discussion of possible modifications of Newtonian
gravity.\cite{talmadge}
But even if more detailed data analysis confirms that
planetary orbits are incompatible with the accelerations acting
on spacecraft, one might still conclude that spacecraft and
planets react differently to some previously unknown
component of the
gravitational interaction.  With this possibility in mind,
more work is being planned on both the theoretical and observational
questions.


\subsection{Planetary scales}

     Anomalies also exist in the Galileo trajectory during the
two close flybys of Earth in December 1990 and December 1992.
The necessity of increasing total orbital energy at the first
flyby in order to fit the radio data, by an equivalent velocity
increase of 4 mm/s, led JPL to schedule tracking of the second
flyby with the Tracking and Data Relay Satellite System (TDRSS)
from earth orbit.\cite{edwards}   After analyzing the
TDRSS data, they were left with systematic effects not removable by
the standard Earth gravity model determined from Earth satellites
(Goddard JGM-2 70X70 gravity
field truncated to a 40X40 field).

Having ruled out the usual
nongravitational forces acting on spacecraft, they currently have
no physical explanation for either anomalous Earth flyby.  Also,
they have tested against software bugs by
successfully comparing results from JPL's Orbit Determination
Program (ODP) with results from Goddard's GEODYNE software.

Similar data anomalies during flybys of other planets are not
detectable because of uncertainties in their gravity fields, with
the possible exception of the Mariner 10 Mercury flyby in March
1975.  There they removed Doppler systematics with an unexpectedly
large gravity anomaly, about one-tenth the largest Earth anomaly
(the antarctic low).  This is large for Mercury but not way
outside the bounds of plausibility.\cite{anderson}
Future Mercury orbiters should tell us whether or not such a
large gravity anomaly is real.


\section{Gravity and Antimatter}

To place the last two sections in perspective, whatever is going on here
 the whole discussion should make it clear to you
that  our understanding of gravity within the universe is
incomplete.

This brings us full circle.  Given that our theoretical and experimental
knowledge of the physics of gravity and antimatter are woefully inadequate,
to perform an experiment on the gravity of antimatter would be a monumental
milestone in our understanding of physics.  This would be true even if we
found exactly what we expect, that gravity on antimatter is the same as that on
matter. Until we actually do such an experiment, we do not know the
answer, we only believe we do.

The proposal to measure the gravitational
acceleration of the antiproton\cite{gn} has progressed to the PS200
experiment.\cite{mhh}  The first part of this experiment is already on the
floor at LEAR,
the ``catching trap."\cite{walt}

There are also two main ideas on how to form
antihydrogen: via positronium-antiproton\cite{deutch} collisions or directly
from positron-antiproton collisions.\cite{gabri}  Then it might be possible to
control this antihydrogen by laser cooling, magnetic traps, or ``fountains."
If so,
a long-term goal would be to measure gravity on antihydrogen.  (See the
discussion in Refs. \cite{walt,trap} for a comparison of these ideas.)

But in any event, it is in the hands of our generation to perform an
experiment to measure the gravitational acceleration of antimatter.  Some day
it will be done, whether we do it or not.  {\it It will be done.} If we do not
do it, and the answer eventually turns out to be what we expect, then future
generations will look back upon us a say it was a shame.  But if the answer
turns out to be a surprise, then, if we do not do it, future generations will
look back upon us and say we were fools.  \\


\noindent {\bf Acknowledgements} \\

There are many individuals who deserve thanks for their help on this
project.  With apologies to those whom I leave out, let me
mention   Jack Hills,  Gabe Luther, and Raymond Laflamme.
JDA and ELL benefitted from many theoretical discussions with
Timothy P. Krisher.  Their research was carried out at the Jet
Propulsion Laboratory, California Institute of Technology, and
was sponsored by the Pioneer Project Office, NASA/Ames Research
Center, LOA PPO-17, through an agreement with the National
Aeronautics and Space Administration.


\end{document}